\documentclass{article}

\usepackage[final]{neurips_2021}

\usepackage[hang,flushmargin]{footmisc}

\usepackage[utf8]{inputenc} 
\usepackage[T1]{fontenc}    
\usepackage[hidelinks]{hyperref}
\usepackage{url}            
\usepackage{booktabs}       
\usepackage{amsfonts}       
\usepackage{nicefrac}       
\usepackage{microtype}      
\usepackage{xcolor}         
\usepackage{amsmath}
\usepackage{amssymb}
\usepackage{enumitem}
\usepackage{graphicx}       

\usepackage{colortbl}
\definecolor{Gray}{gray}{0.5}
\definecolor{GrayBG}{gray}{0.95}

\title{Sparsifying Sparse Representations for Passage Retrieval by Top-$k$ Masking}

\author{
  {\bf Jheng-Hong Yang, Xueguang Ma, and Jimmy Lin} \\[1ex]
  David R. Cheriton School of Computer Science \\
  University of Waterloo
}

\begin{document}

\maketitle

\begin{abstract}
Sparse lexical representation learning has demonstrated much progress in improving passage retrieval effectiveness in recent models such as DeepImpact, uniCOIL, and SPLADE.
This paper describes a straightforward yet effective approach for sparsifying lexical representations for passage retrieval, building on SPLADE by introducing a top-$k$ masking scheme to control sparsity and a self-learning method to coax masked representations to mimic unmasked representations.
A basic implementation of our model is competitive with more sophisticated approaches and achieves a good balance between effectiveness and efficiency.
The simplicity of our methods opens the door for future explorations in lexical representation learning for passage retrieval.
\end{abstract}

\section{Introduction}

Representation learning for text ranking using pretrained transformer-based language models have dramatically improved the effectiveness of passage retrieval in recent years.
There are two main approaches:

\begin{itemize}
\item {\bf dense representations}~\citep{omar2020colbert,karpukhin-etal-2020-dense,xiong2020approximate,sebastian2021tasb,lin-etal-2021-batch}, and
\item {\bf sparse representations}~\citep{zamani2018neural,dai2020ct,bai2020sparterm,mallia2021di,lin2021brief,zhuang2021fast,formal2021spladev1,formal2021spladev2}.
\end{itemize}

The former class aims to produce dense low-dimensional vector representations while the latter class aims to produce sparse high-dimensional vector representations (with few non-zero elements).
In the conceptual framework proposed by~\cite{lin2021proposed}, this design choice is a component of the logical scoring model that determines how query--document relevance scores are computed.
Searching over these types of representations (i.e., top-$k$ retrieval) is determined by different physical retrieval models.
For example, searching dense representations is a nearest neighbor search problem, typically performed using libraries such as Faiss~\citep{JDH17} and NMSLib~\citep{boytsov2013nms}.
In contrast, sparse representations typically use inverted indexes, with toolkits such as Anserini (based on Lucene)~\citep{Yang_etal_JDIQ2018}, PISA~\citep{mallia2019pisa}, and JASS~\citep{lin2015anytime,trotman2019micro}.

We argue that ``dense representations'' are perhaps better called latent representations, because the basis of the vector space derives from the latent features captured in pretrained language models such as BERT.
Correspondingly, ``sparse representations'' should be called lexical representations, because the dimensions of their vectors derive from tokens in the vocabulary.
In other words, the salient distinction is the basis of the vector space, not density or sparsity per se.
For example, it is actually possible to take DPR ``dense'' representations and sparsify them~\citep{yamada-etal-2021-efficient}, and as we show later, a ``sparse'' model such as SPLADE can actually generate lexical representations that are quite ``dense'' in having many non-zero elements.
That is, lexical representations can either be dense or sparse.
Accordingly, we adopt this terminology and refer to our work as learning sparse lexical representations.
The weights in our vectors can be interpreted as impact scores~\citep{Anh_etal_SIGIR2001} and top-$k$ retrieval can be performed using standard inverted indexes.

In this work, we build on the SParse Lexical AnD Expansion (SPLADE) model proposed by~\cite{formal2021spladev1,formal2021spladev2} and add two innovations:\
The first is to incorporate a top-$k$ masking scheme to sparsify the high-dimensional and dense representations generated by the MLM projection head.
In addition, we explore a self-learning method inspired by the work of~\cite{gao-etal-2021-simcse} to coax the top-$k$ masked representations to mimic their unmasked counterparts.
These two proposed additions greatly improve effectiveness on the MS MARCO passage ranking test collection~\citep{MS_MARCO_v3}.
Our improved model, which we call SPLADE-mask, is competitive with state-of-the-art lexical retrieval models that incorporate more complex multi-stage training regimes~\citep{mallia2021di,lin2021brief,zhuang2021fast}, hard negative mining, or knowledge distillation using pretrained cross-encoders~\citep{formal2021spladev2}.

\section{Preliminaries}

Before elaborating on the details of our method, we first present some necessary notation.
Our method is based on the logical scoring model proposed by~\cite{lin2021proposed}, which defines the query--document scores for document ({\it ad hoc}) retrieval.
Given a query $q$ and a document $d$ with their fixed-width vector representations $\textrm{\bf q}$ and $\textrm{\bf d}$, which are generated by two arbitrary functions $\eta_{q}(\cdot)$ and $\eta_d(\cdot)$, a logical scoring model defines the degree $s$ to which $d$ is relevant to $q$ as:
\begin{equation*}
    s(q, d) \triangleq \phi(\eta_q(q), \eta_d(d)) = \phi(\textrm{\bf q}, \textrm{\bf d}),
\end{equation*} 
where $\phi$ is a similarity operator defined in a metric space.
Today, $\eta_{q}(\cdot)$ and $\eta_d(\cdot)$ are most commonly implemented using transformed-based pretrained language models.

\paragraph{Lexical representation learning for text ranking.}
Lexical representation learning describes a number of related models for using pretrained language models to predict term importance~\citep{dai2020ct,bai2020sparterm,mallia2021di,lin2021brief,zhuang2021fast,formal2021spladev1,formal2021spladev2}.
At a high level, these models learn to assign term weights in a predefined vocabulary space, usually composed of the tokens used in pretrained language models.
That is, the dimensions of $\textrm{\bf q}$ and $\textrm{\bf d}$ are defined by the vocabulary.
The learning objective is to discriminate representations of relevant vs.\ non-relevant pairs based on training data.
A representative loss function for this task is:
\begin{equation}
\label{eq:infonce}
    \mathcal{L}_\text{rank} = - \log \frac{e^{\phi(\textrm{\bf q}, \textrm{\bf d}_{+})}}{e^{\phi(\textrm{\bf q}, \textrm{\bf d}_{+})} + \sum_{\textrm{\bf d}_{-}} e^{\phi(\textrm{\bf q}, \textrm{\bf d}_{-})}}.
\end{equation}
\noindent Here, $\textrm{\bf q}$ is the query representation, $\textrm{\bf d}_{+}$ is the representation of a positive (relevant) document, $\{\textrm{\bf d}_{-}\}$ are representations of negative (non-relevant) documents, and $\phi$ is a similarity measure, e.g., inner product or cosine similarity.
Typically, a positive pair comprising a query and a relevant document is provided as part of the training data.
Negative documents can be sampled from the corpus using a unsupervised lexical matching model like BM25~\citep{karpukhin-etal-2020-dense}, a previous version of the text ranking model itself~\citep{xiong2020approximate}, or in-batch negatives using relevant documents associated with other queries~\citep{henderson2017efficient}.

In lexical representation learning, query and document representations are projected into a vocabulary space whose dimension is defined by the number of predefined tokens $|V|$.
For example, suppose lexical representation learning is based on BERT.
In this case, the query and document representations are formed by vectors of $|V| = 30522$ dimension, where the WordPiece tokens serve as the basis of the vector space.
Eq.~(\ref{eq:infonce}) is minimized by learning the term weights of the $|V|$-dimensional representations of queries and documents.

\paragraph{Sparsifying lexical representations.}
The $|V|$-dimensional lexical representations of queries and documents produced by $\eta_{q}(\cdot)$ and $\eta_d(\cdot)$ are not guaranteed to be sparse vectors, i.e., there might be many non-zero elements.
Therefore, existing works use different approaches to regularize the representation space when optimizing Eq.~(\ref{eq:infonce}).
For instance, DeepImpact~\citep{mallia2021di} and uniCOIL~\citep{lin2021brief} only assign non-zero weights to input tokens of queries and documents (with document expansion as a preprocessing step) instead of using the full $|V|$-dimensional space.
In this paper, we explore sparse lexical representations based on the recently proposed SPLADE model~\citep{formal2021spladev1,formal2021spladev2} that utilizes the full $|V|$-dimensional space but applies a sparsity constraint.

Formally speaking, given a query or a document, let us consider an input token sequence of length $N$ and its contextualized representations $\{\textrm{\bf c}_i\}_{i=1}^{N}$ generated by a BERT model, where $\textrm{\bf c} \in \mathbb{R}^{h}$; $h$ is the dimension of the hidden layer.
The $|V|$-dimensional representation $\{w_{i,j}\}_{j=1}^{|V|}$ is produced by the MLM head on top of ${\textrm{\bf c}_i}$, and the predicted term weight at the $j$-th dimension is given by:
\begin{equation*}
    w_{i,j} = \psi(\textrm{\bf c}_i)^{T}\textrm{\bf e}_{j} + b_{j},
\end{equation*}
\noindent where $\psi$ is a transformation composed of a linear layer with GeLU activation and LayerNorm operation, $\textrm{\bf e}_{j}$ is the $j$-th row of the vocabulary projection matrix, and $b_j$ is the bias parameter.
The final representation of the input sequence is then obtained by conducting a pooling operation on the set of $|V|$-dimensional representations with $N$ elements.
For instance, in SPLADEv2~\citep{formal2021spladev2}, the $j$-th weight is defined as:
\begin{equation}
\label{eq:splade}
    w_{j} = \max_{i} \log \left(1 + \text{ReLU}(w_{i,j}) \right).
\end{equation}
\noindent Note that this design already imposes an implicit sparsity constraint on the lexical representations by using a combination of the ReLU activation function and pretrained vocabulary projection matrix $\{\textrm{\bf e}_j\}_{j=1}^{|V|}$.
In order to further sparsify the final lexical representations, \cite{formal2021spladev1} add a FLOPS regularization loss~\citep{paria2020minimizing}.

\begin{figure}[t]
\centering
\includegraphics[width=\textwidth]{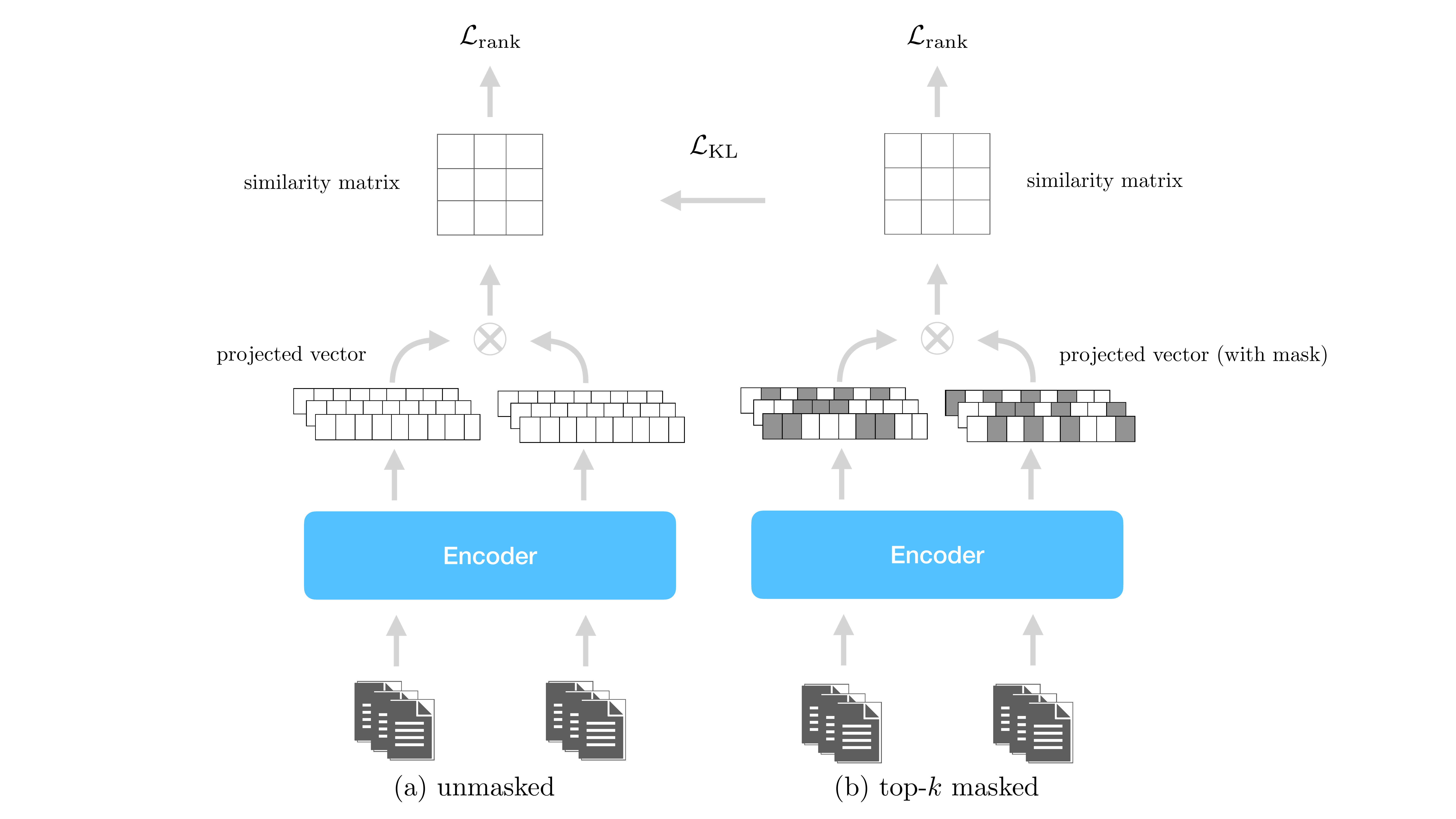}
\caption{Overview of our approach to learning lexical representations for passage retrieval.
Queries and passages are encoded and projected into a high dimensional representation space (defined in terms of the transformer vocabulary), where pairwise similarities computed via inner products form the basis for optimization based on training data.
Without any sparsity constraint, the learned representations may be dense (i.e., have many non-zero elements), as shown on the left.
We propose to use top-$k$ masking to control the sparsity of the projected representations (shown on the right) and further coax sparse representations to mimic unmasked representations.}
\label{fig:teaser}
\end{figure}

\section{Approach}
\label{section:approach}

Building on SPLADE, we propose a simple yet effective method to control the sparsity of the final lexical representation.
Specifically, our motivation is to learn lexical representations with precise control of the portion of activated elements (i.e., dimensions or tokens that are assigned non-zero weights).
By doing so, we can directly optimize the lexical representations for scenarios with different storage requirements.
Since the ReLU operation already ensures that term weights are positive, a natural extension is to impose a top-$k$ masking scheme on top of Eq.~(\ref{eq:splade}).
Specifically, we only keep the top-$k$ activated weights in our final representations:
\begin{equation}\label{eq:topk}
    w_j' = \max_{j}{}_{k}\{ w_j\}_{j=1}^{|V|},
\end{equation}
\noindent where we define the operation $\max{}_{k}$ as keeping the top-$k$ activated weights and setting the other dimensions to zero.
There are two possible implementations:
\begin{enumerate}[leftmargin=*]
    \item Post processing: apply top-$k$ masking in the inference stage after training is complete, or
    \item Joint training: apply top-$k$ masking in both stages.
\end{enumerate}
\noindent When applying the joint training approach, we can set $k$ to a constant (i.e., a hyperparameter) or change it gradually with an exponential decay scheduler:
\begin{equation}
k' = |V| \times (1 - r_{\text{decay}})^{\frac{\text{steps}}{\text{steps per epoch}}},
\end{equation}
where $r_{\text{decay}}$ is a predefined decay rate (i.e., another hyperparameter).

One additional benefit of our proposed masking scheme is that we can incorporate a self-learning method with joint training to improve ranking effectiveness, as shown in Fig.~\ref{fig:teaser}.
Inspired by the work of~\cite{gao-etal-2021-simcse}, we add a regularization term to coax the top-$k$ activated weights produced by the masking scheme to mimic the unmasked ones.
Specifically, we use KL-divergence as follows:
\begin{equation}
\label{eq:kl}
    \mathcal{L}_\text{KL} = D_{\text{KL}}(P_{\text{top-}k} || P_{\text{unmasked}}),
\end{equation}
\noindent where $P$ is defined as softmax normalized probabilities produced by the pairwise similarities (measured by inner product) of a query representation $\textrm{\bf q}$ and its positive and negative document representations ($\textrm{\bf d}_{+}$ and $\textrm{\bf d}_{-}$, respectively).

\section{Experiments}

\paragraph{Settings.}
Our experiments are conducted on the MS MARCO passage corpus~\citep{MS_MARCO_v3}.
We follow two standard evaluation protocols for the passage ranking task:

\begin{enumerate}[leftmargin=*]
\item MS MARCO Dev:\ we report MRR@10 using the sparse binary judgments of the MS MARCO dev subset, comprising 6980 queries;
\item TREC 2019 \& 2020 Deep Learning (DL) Tracks:\ for the TREC DL 2019 and 2020 evaluation sets with dense graded judgments, we report nDCG@10, the official metric.
\end{enumerate}

\noindent As for the training data construction, we use the same settings (except where noted otherwise) as the codebase of Tevatron,\footnote{\url{https://github.com/texttron/tevatron/tree/main/examples/msmarco-passage-ranking}} a toolkit for training bi-encoder retrieval models.
Specifically, we only use labeled positives and BM25 negatives to construct our training triples.
When optimizing the objectives of Eq.~(\ref{eq:infonce}) and Eq.~(\ref{eq:kl}), we consider both BM25 negatives and in-batch negatives.
All of our models use \texttt{distilbert-base-uncased} as the backbone encoder~\citep{sanh2019distilbert}.
The hyperparameters in our reproduced SPLADE model are also used in the other variants.
We train our models with a batch size of 48 and a learning rate of $10^{-5}$ for 20 epochs using the AdamW optimizer.

Finally, we use Anserini~\citep{Yang_etal_JDIQ2018} to construct an inverted index for retrieval, where the token scores are quantized by multiplying the floating point weights by $10^2$ and then rounding to the nearest integer.
This essentially creates impact weights for each document in the corpus and retrieval becomes the sum of impact scores~\citep{Anh_etal_SIGIR2001} of matching query terms; Anserini has implementations for this use case.
For fair comparisons, we report the sizes of the underlying Lucene inverted indexes built by Anserini for all models.

\begin{table}[t]
\begin{center}
\resizebox{\textwidth}{!}{
\begin{tabular}{l|ccccc|rrr}
~  & \multicolumn{5}{c|}{Training method} & \multicolumn{3}{c}{MS MARCO Passage} \\
case &
flops &
freeze &
top-$k$ &
decay &
kl &
Dev MRR@10 & 
Storage (GiB) &
Avg. tokens\\
\toprule
\color{Gray}{(1) SPLADE-max~\citep{formal2021spladev2}} & \color{Gray}{\checkmark} & & & & & \color{Gray}{0.340} & \color{Gray}{2.0} & 96.7\\
(2) SPLADE-max(reproduced) & & & & & & 0.354 & 30.0 & 1953.1\\
\midrule
(a) & & \checkmark & & & & 0.361 & 35.8 & 2686.9\\
\rowcolor{GrayBG} (b) SPLADE-mask-base & & \checkmark & \checkmark & & & 0.351 & 5.5 & 305.0\\
(c) &  & \checkmark & \checkmark & & & 0.349 & 5.4 & 305.0\\
(d) &  & \checkmark & \checkmark & \checkmark & & 0.345 & 5.5 & 305.0\\
(e) &  & \checkmark & \checkmark &  & \checkmark & 0.368 & 5.5 & 305.0\\
\rowcolor{GrayBG} (f) SPLADE-mask &  & \checkmark & \checkmark & \checkmark & \checkmark & 0.373 & 5.4 & 305.0\\
\end{tabular}
}
\end{center}
\vspace{0.15cm}
\caption{Experimental evaluation of our top-$k$ masking scheme and self-learning method using \texttt{distilbert-base-uncased} as the backbone on the MS MARCO passage ranking test collection.
The final column shows the average number of tokens assigned non-zero weights per passage, which is a measure of representation sparsity.
``flops'': with FLOPS regularization;
``freeze'': with frozen MLM token projection layer;
``top-$k$'': with our top-$k$ masking scheme;
``decay'': with an exponential decay schedule to decrease $k$;
``kl'': with our self-learning method to coax the top-$k$ masked representations to mimic the unmasked ones with the KL-divergence objective.
}
\label{tab:ablation}
\end{table}

\paragraph{Sparsifying lexical representations.}
As previously discussed, our work builds on SPLADEv2~\citep{formal2021spladev2}, which uses the Masked Language Model (MLM) token projection head and max pooling to produce lexical representations for queries and passages using DistilBERT.
In~\citet{formal2021spladev2}, the authors further sparsified the output representations by adding FLOPS regularization~\citep{paria2020minimizing} during the training process.
Here we aim to explore other options for sparsifying the representations produced by the MLM head:\ top-$k$ masking and our self-learning method.
We report experimental results in Table~\ref{tab:ablation}.

We first reproduce SPLADEv2 to explore its best possible effectiveness.
Row (1) shows figures on MS MARCO dev copied from the original paper~\citep{formal2021spladev2}.
Without FLOPS regularization, the ranking effectiveness of our reproduced SPLADE-max model improves from 0.340 to 0.354, shown in row (2), but this also increases the index size from 2.0 to 30.0 GiB.
The reason for this can be found in the ``Avg.\ tokens'' column, which shows the average number of tokens with non-zero weights in each passage vector as a measure of sparsity.
We observe a many-fold increase from row (1) to row (2).

In our next experiment, we freeze the linear vocabulary projection layer in the MLM head.
Our intuition is as follows:\ we suspect that the vocabulary weight matrix is already well trained from the MLM pretraining phase, and further adjustments during representation learning for text ranking might incorporate too much bias because it is the last layer of the deep neural network.
This intuition is indeed borne out experimentally, as ranking effectiveness further improves to 0.361, but with a larger index size 35.8 GiB; this is shown in row (a) in Table~\ref{tab:ablation}.

Building on the experiments above, we then explore the techniques discussed in Section~\ref{section:approach}, whose results are shown in rows (b)--(f).
Our first attempt to reduce the index size is to put a top-$k$ mask on the representations, since the weights of SPLADE-max are always greater than zero.
We set $k = 0.01 \times |V| \approx 305$ for DistilBERT.
As shown in row (b) as SPLADE-mask-base, this masking method yields good ranking effectiveness, at 0.351, but with a much smaller index size:\ 5.5 GiB.
We emphasize here that SPLADE-mask-base does not involve any joint training; we're simply ``post processing'' the outputs of row (a).

Interestingly, Table~\ref{tab:ablation} shows that joint training with top-$k$ masking performs slightly worse than SPLADE-mask-base, shown in row (c), which uses a constant $k$ schedule, and row (d), which uses the exponential decay scheduler with $r_{\text{decay}} = 0.2$.
It is not until when we incorporate self-learning via the KL-divergence loss, shown in row (e), that joint training actually improves ranking effectiveness, to 0.368.
Finally, we further improve ranking effectiveness with the combination of top-$k$ masking, exponential decay scheduler for $k$ (same as row (d)), and our self-learning method via KL-divergence loss.
With this full model, which we call SPLADE-mask, we reach 0.373 MRR@10 on the MS MARCO dev queries, as shown in row (f).

\begin{table*}[t]
\centering
\resizebox{\textwidth}{!}{
\begin{tabular}{lllrrrr}
\toprule
\multicolumn{3}{c}{} & Dev & TREC DL19 & TREC DL20 & Storage \\
\multicolumn{3}{c}{Model} & MRR@10 & nDCG@10 & nDCG@10 & (GiB) \\
\toprule
\multicolumn{7}{c}{\emph{direct weighting methods}} \\
\midrule
(a) \citet{Yang_etal_JDIQ2018} & BM25          &            & 0.184 & 0.506 & 0.480 & 0.6  \\
(b) \citet{dai2020ct}          & DeepCT        &            & 0.243 &      &      &      \\
(c) \citet{formal2021spladev2} & SPLADE-max    &            & 0.340 & 0.684 &      & 2.0  \\
(d) \citet{formal2021spladev2} & SPLADE-distil  & & 0.368 & 0.729 & 0.711 & 5.0 \\
\rowcolor{GrayBG} (e) \textbf{This work} & \textbf{SPLADE-mask}  & & \textbf{0.373} & 0.707 & 0.678 & 5.4 \\
\midrule
\multicolumn{7}{c}{\emph{with explicit document expansion}} \\
\midrule
 & Scoring & Expansion & \\
\cmidrule(lr){2-2} \cmidrule(lr){3-3}
(f) \citet{rodrigo2019dt5q}    & BM25          & doc2query-T5 & 0.277 & 0.648 & 0.616 & 1.0  \\
(g) \citet{mallia2021di}       & DeepImpact    & doc2query-T5 & 0.325 &      &      & 1.4  \\
(h) \citet{lin2021brief}       & uniCOIL       & doc2query-T5 & 0.351 & 0.693 & 0.666 & 1.3  \\
(i) \citet{zhuang2021fast}     & uniCOIL       & TILDE      & 0.350 & 0.728 & 0.711 & 2.1  \\
\bottomrule
\end{tabular}
}
\caption{The effectiveness of various retrieval models using lexical representations on the MS MARCO passage corpus.
}
\label{tab:main}
\end{table*}

\paragraph{Comparison with other lexical retrieval models.}
Table~\ref{tab:main} compares several recent sparse lexical retrieval models with our best model configuration, the SPLADE-mask variant shown in Table~\ref{tab:ablation}(f).
Also, we report results on TREC DL19 and DL20 to examine the effectiveness of our model on dense graded relevance judgments.

We can identify two categories of models, shown as separate blocks in Table~\ref{tab:main}.
In the first block, rows (a)--(e), we list models that directly encode term weights, with BM25 as the baseline, shown in (a).
In the second block, we report models that apply a more complex training regime that first incorporates an explicit document expansion phase; these are reported in rows (f)--(i).
Row (f), which applies document expansion using doc2query-T5~\citep{rodrigo2019dt5q} but retains BM25 term weighting, can be viewed as a baseline.
Our SPLADE-mask model can be viewed as an instance of the first category, and in general, we find that these methods appear to be more effective.

Row (d) reports the results of SPLADE-distil~\citep{formal2021spladev2}, which represents to our knowledge the state of the art in lexical representation learning.
This model builds on the basic SPLADE design, but further incorporates training with hard negatives and cross-encoder distillation.
Our approach achieves the highest ranking effectiveness on MS MARCO, reaching 0.373 MRR@10 with an index size of 5.4 GiB.
Note that we are able to beat SPLADE-distil with a rather simple ``base'' design---that is, we have not incorporated hard negatives and cross-encoder distillation.
We believe that those features are orthogonal to the techniques proposed here, and incorporating them into SPLADE-mask might further boost effectiveness.
We leave this for future work.

In terms of index size, our approach consumes slightly more space than SPLADE-distil and much more space than methods with explicit document expansion.
For example, uniCOIL with doc2query-T5 expansion~\citep{lin2021brief} is two points worse on MS MARCO dev but its index size is four times smaller. 
On TREC DL19 and DL20, the effectiveness of our SPLADE-mask model is lower than the  other top models, but we currently have no explanation for this.

\section{Future Work and Conclusions}

Putting everything together, the two relatively simple ideas we propose in this paper---top-$k$ masking and self-learning---appear to yield noticeable increases in retrieval effectiveness.
Furthermore, these two innovations can be further integrated with other features of the full model described by \citet{formal2021spladev2}, hard negatives and cross-encoder distillation, which can further boost effectiveness.
We have shown that SPLADE-mask achieves a good balance between effectiveness and index size, but we have yet to examine query latency as an important consideration.
In this regard, \cite{Mackenzie_etal_arXiv2021} has shown that the SPLADE family of models is still quite a bit slower than, for example, uniCOIL.
More in-depth analyses of effectiveness--efficiency tradeoffs are needed, but we are excited about future developments of learned lexical representations for retrieval.

\section*{Acknowledgements} 

This research was supported in part by the Canada First Research Excellence Fund and the Natural Sciences and Engineering Research Council (NSERC) of Canada.

\bibliographystyle{abbrvnat}
\bibliography{ref}

\end{document}